\documentclass[aps, prl, preprint, groupedaddress, superscriptaddress]{revtex4-1}

\usepackage{graphicx}   
\usepackage{amsmath}   
\usepackage[utf8]{inputenc}
\usepackage{siunitx}

\begin{document}

\title{Experimental determination of exchange constants in antiferromagnetic $\mathbf{Mn_2Au}$}

\author{A.A.\,Sapozhnik}
\affiliation{Institut f\"ur Physik, JGU Mainz, Staudingerweg 7, 55128, Mainz, Germany}
\affiliation{Graduate School Materials Science in Mainz, Staudingerweg 9, 55128, Mainz, Germany}
\author{C.\,Luo}
\affiliation{Helmholtz-Zentrum Berlin f\"ur Materialien und Energie, Albert-Einstein Str. 15, 12489, Berlin, Germany}
\affiliation{Institut für Experimentelle und Angewandte Physik, Universität Regensburg, Universitätsstrasse 31, 93053 Regensburg, Germany}
\author{H.\,Ryll}
\affiliation{Helmholtz-Zentrum Berlin f\"ur Materialien und Energie, Albert-Einstein Str. 15, 12489, Berlin, Germany}
\author{F.\,Radu}
\affiliation{Helmholtz-Zentrum Berlin f\"ur Materialien und Energie, Albert-Einstein Str. 15, 12489, Berlin, Germany}
\author{M.\,Jourdan}
\affiliation{Institut f\"ur Physik, JGU Mainz, Staudingerweg 7, 55128, Mainz, Germany}
\affiliation{Graduate School Materials Science in Mainz, Staudingerweg 9, 55128, Mainz, Germany}
\author{H.\,Zabel}
\affiliation{Institut f\"ur Physik, JGU Mainz, Staudingerweg 7, 55128, Mainz, Germany}
\affiliation{Graduate School Materials Science in Mainz, Staudingerweg 9, 55128, Mainz, Germany}
\author{H.-J.\,Elmers}
\affiliation{Institut f\"ur Physik, JGU Mainz, Staudingerweg 7, 55128, Mainz, Germany}
\affiliation{Graduate School Materials Science in Mainz, Staudingerweg 9, 55128, Mainz, Germany}

\begin{abstract}

Mn$_2$Au is an important antiferromagnetic (AF) material for spintronics applications. Due to its very high Néel temperature of about \SI{1500}{\kelvin}, some of the basic properties are difficult to explore, such as the AF susceptibility and the exchange constants. Experimental determination of these properties is further complicated in thin films by unavoidable presence of uncompensated and quasi-loose spins on antisites and at interfaces. Using x-ray magnetic circular dichroism (XMCD), we have measured the spin and orbital contribution to the susceptibility in the direction perpendicular to the in-plane magnetic moments of a  Mn$_2$Au(001) film and in fields up to $\pm$\SI{8}{\tesla}. By performing these measurements at a low temperature of \SI{7}{\kelvin} and at room temperature, we were able to separate the loose spin contribution from the susceptibility of AF coupled spins. The value of the AF exchange constant obtained with this method for a \SI{10}{\nano\metre} thick Mn$_2$Au(001) film equals to \SI[separate-uncertainty=true]{24(5)}{\milli\electronvolt}. 

\end{abstract}


\maketitle

\section{I. INTRODUCTION}

At present, antiferromagnets (AF) play an important role in the rapidly growing field of AF spintronics \cite{BAL17} due to their fast THz dynamics \cite{KIM04, KAM10, BOS16, BOW16}. In the past, AF materials have been exploited for providing exchange bias in spin valves \cite{NOG99}. However, at present AF materials are in the focus of interest for their own right, for instance, for magnetization switching by (unpolarized) electrical currents, proposed for designing the next generation data storage devices \cite{WAD16, BOD18}. One of the main requirements for prospective AF applications is the magnetic order stability at room temperature. The characteristic temperature up to which the material preserves its magnetic order is called Néel temperature ($T_N$). The majority of the antiferromagnetic materials have $T_N$ either lower or close to room temperature \cite{BAL17}. Only a few metallic AF materials exist, which are suitable for spintronics application above room temperature, including Mn$_2$Au with a very high $T_N$ of \SI{1500}{\kelvin} in a bulk crystal \cite{BAR13}.

According to quantum statistics considerations, $T_N$ depends on the exchange interaction strength in a material, which defines the mean effective Weiss field stabilizing the antiparallel spin arrangement. The exchange interaction between two spin moments $i^{th}$ and $j^{th}$ can be described in terms of exchange constants $J_{ij}$. The value of the $J_{ij}$, in turn, depends on the distance between corresponding atoms and decreases exponentially with increasing distance \cite{GET08}. Another characteristic of non-perfect AF materials is the existence of uncompensated "loose" spins, due to antisites and disorder at interfaces. Loose spins play an important role in exchange biased ferromagnet/AF bilayers, since it is believed that they are responsible for broadening of the hysteresis loop \cite{RAD08}. In addition, loose spins also occur at grain boundaries in polycrystalline and epitaxial thin films. 

The exchange constants $J_{ij}$ of antiferromagnets can be determined from the perpendicular susceptibility, i.e. the spin response to a magnetic field applied perpendicular to the AF spins. The standard tools for measuring the perpendicular susceptibility are either magnetometry or neutron scattering. In contrast to the mentioned methods, x-ray magnetic circular dichroism (XMCD) provides element specific data about the magnetic properties of the investigated material and is sensitive to several nanometers at the surface in the total electron yield (TEY) mode. Owing to the sum rules \cite{THO92, CAR93}, XMCD allows accessing the spin and orbital magnetic moment values separately, leading to further insight into the magnetic properties of AFs.

In this contribution, we report on experimental results for the perpendicular magnetic susceptibility in the ordered AF phase of Mn$_2$Au using XMCD. The results obtained at room temperature and at \SI{7}{\kelvin} allow to distinguish contributions from weakly coupled (loose) spins with paramagnetic behavior and AF coupled spins with small constant perpendicular susceptibility. From the response of AF coupled spins we obtain an experimental value for the exchange constant. In addition, the ratio of the measured orbital and spin magnetic moments is compared to our density functional theory (DFT) calculations and sheds light on the AF moment coupling mechanism.

\section{II. METHODS}

Mn$_2$Au epitaxial thin films were deposited with the rf-magnetron sputtering technique. The base pressure in the sputtering chamber was $\sim$\SI{e-8}{\milli bar}. The sample was prepared with the following layer structure Al$_2$O$_3$(1$\bar{1}$02) substrate/Ta(001) \SI{15}{\nano\metre}/Mn$_2$Au(001) \SI{10}{\nano\metre}/AlO$_x$ \SI{2}{\nano\metre}. The Ta buffer ensures the necessary (001) orientation of Mn$_2$Au. The film surface was coated with a \SI{2}{\nano\metre} thick Al film for oxidation protection. The sample structure was characterized with X-ray diffraction (XRD) and reflection high-energy electron diffraction (RHEED) to prove its high surface and bulk quality \cite{JOU15}. The average grain size in the Mn$_2$Au film estimated from the XRD spectrum using the Scherrer equation is \SI{15}{\nano\metre} in the sample plane. A stoichiometry of \SI[separate-uncertainty=true]{66.2(3)}{\percent} Mn and \SI[separate-uncertainty=true]{33.8(3)}{\percent} Au was determined by energy-dispersive X-ray spectroscopy (EDX) indicating a slight Au excess. Details about the preparation and characterization procedures are reported in \cite{JOU15}.

The XMCD spectroscopy measurements were performed in the VEKMAG end station \cite{NOL16} at BESSY II (HZB, Berlin) during the multi-bunch operational mode of the synchrotron storage ring. The sample was placed in the chamber at a pressure of $\sim$\SI{2e-10}{\milli bar} in the gap of a superconducting magnet providing a maximum of $\pm$\SI{9}{\tesla} along the beam direction. The photocurrent extracted from the sample reached several picoamperes. The X-ray absorption spectra (XAS) were measured for X-ray energies close to the $L_3$ and $L_2$ absorption edges of Mn. The photocurrent was normalized to the current of a Pt-mesh monitor located in the X-ray beam path in order to compensate for variations of the ring current and for the transmission function of the optical elements. The XMCD spectrum was obtained as the difference of XAS data measured for circular right and left X-ray polarization with the polarization of \SI{77}{\percent}, for which the largest figure of merit is reached.

The Mn$_2$Au density of states (DOS) was calculated with the FP-LAPW ELK software \cite{ELK}. The lattice constants were set equal to the published values of $a$ = \SI{0.3328}{\nano\metre} and $c$ = \SI{0.8539}{\nano\metre} \cite{KHM08}. The staggered magnetization direction orientation was chosen to be along the $\langle 110 \rangle$ crystallographic directions \cite{BAR15}. The spin-orbit coupling is self-consistently added to the second-variational Hamiltonian. A set of 1728 k-points distributed over the first Brillouin zone was used in the numeric scheme. The total energy error was less than \SI{0.1}{\milli\electronvolt}/f.u..

\section{III. THEORETICAL RESULTS}

\begin{figure}[ht]
\includegraphics[scale=0.35]{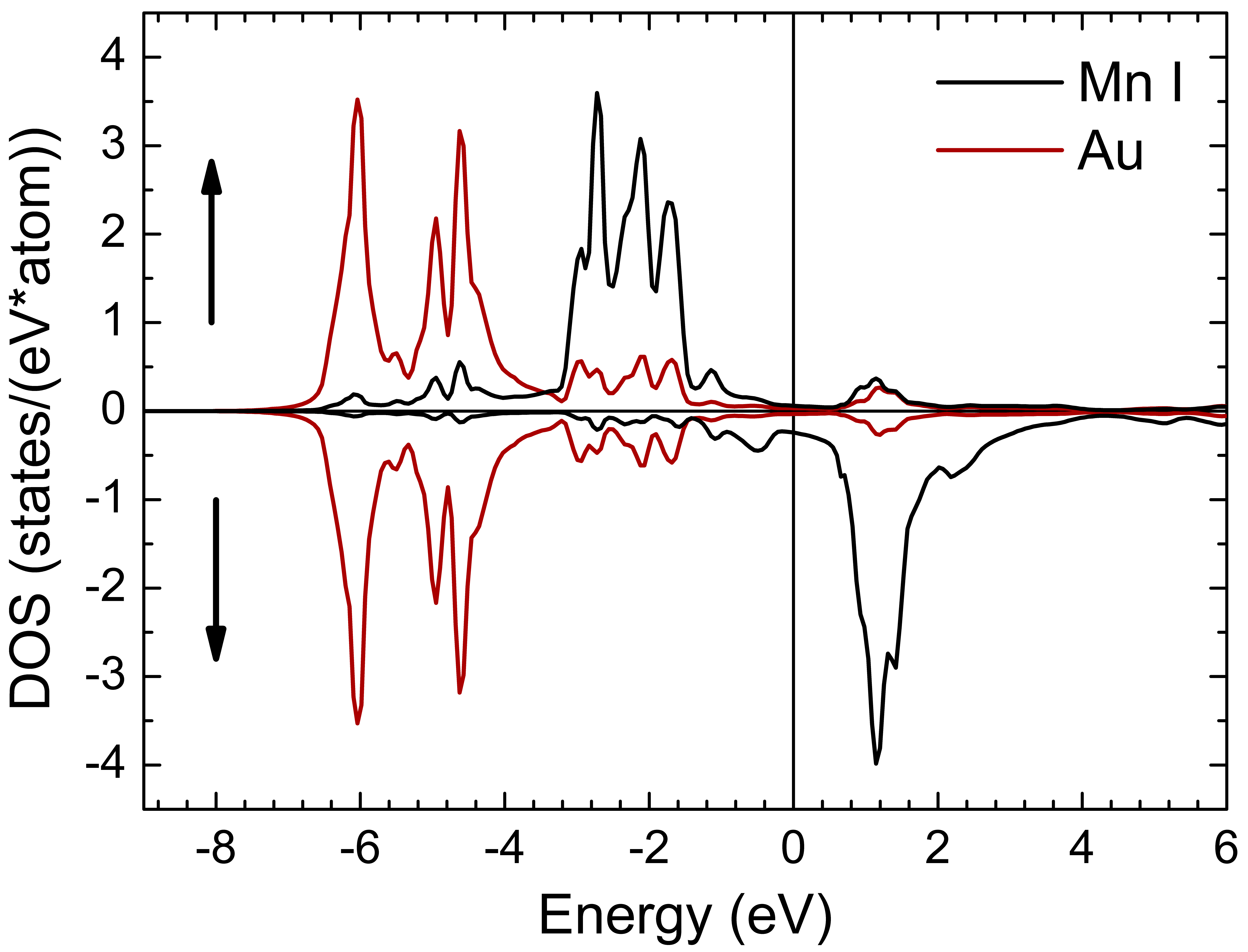}
\caption{Element resolved DOS of Mn$_2$Au. Positive and negative DOS values correspond to spin-up (majority) and spin-down (minority) electrons, respectively.}
\label{fig:MADOS} 
\end{figure}

It was demonstrated by Barthem et al. \cite{BAR13} that Mn$_2$Au has a layered antiferromagnetic structure with two sublattices denoted as Mn\,I and Mn\,II. The calculated Mn$_2$Au DOS indicates that the Mn atoms carry a magnetic moment (Fig.\,\ref{fig:MADOS}), whereas the Au atoms do not possess a magnetic moment, confirmed by the identical DOS for majority and minority electrons. The DOS of Mn\,II atoms (not shown) is found to be opposite to the DOS of Mn\,I atoms.

The simulation provides the following Mn spin moment $\mu_S^{Mn} = \pm3.591\,\mu_B$ and orbital moment $\mu_L^{Mn} = \mp0.002\,\mu_B$. The value of $\mu_L^{Mn}$ is much smaller than $\mu_S^{Mn}$, which is typical for transition metal elements due to orbital moment quenching \cite{GAL05}. Another important property of Mn$_2$Au obtained from the DOS is the number of d-holes per Mn atom $N_d$, which amounts to 5.0.

In the following, we discuss expected contributions to the field-induced magnetic moments based on the calculated total magnetic moments. In general, the perpendicular magnetization induced in Mn$_2$Au by an external field contains two contributions. One of them is generated by the rotation of the exchange coupled moments. Due to the high $T_N$ of Mn$_2$Au, thermal excitations can be neglected below room temperature. Thus, the exchange torque contribution is expected to be proportional to the applied field. The ratio of the resulting magnetic moment in the direction perpendicular to the film and the applied magnetic field value then provides the perpendicular susceptibility $\chi_{\perp} = M_{\perp}/(\mu_0 H)$. In this case, the induced magnetic moment is usually very small. The order of magnitude of the induced moment for Mn$_2$Au can be estimated from the susceptibility $\chi$\,=\,\num{5e-4} measured in Ref.\,\cite{BAR13}. Using the saturation magnetization of Mn$_2$Au $M_S$\,=\,\SI{1.56e6}{\ampere\per\metre} results in $M_{\perp}/M_S$\,=\,\SI{0.2}{\percent} at \SI{8}{\tesla}. However, the powder samples investigated in Ref.\,\cite{BAR13} represent an average over randomly oriented bulk crystals. The second contribution originates from weakly coupled or loose Mn moments. Their behavior resembles a paramagnet and can be described by the Brillouin function:

\begin{equation}
\frac{M^{loose}}{M^{loose}_S} = B_J\left(\frac{g \mu_B J \mu_0 H}{k_B T}\right),
\label{eq:Brillouin}
\end{equation}

where $M^{loose}$ and $M^{loose}_S$ are the induced perpendicular magnetization and saturation magnetizations of the loose spins, respectively, and $g$ is the gyromagnetic ratio. Since $\mu_L^{Mn}$ is much smaller than $\mu_S^{Mn}$, $J$ is replaced by $S=1.8$ in the following discussion. $M^{loose}_S$ depends on the concentration of the uncoupled spins, the value of which is to be found from the experiment. However, in epitaxial magnetic films their concentration is expected to be rather small. This assumption is scrutinized for our samples in Section~V. Substituting the experimental values of $\mu_0 H$ = \SI{8}{\tesla} and $T$ = \SI{300}{\kelvin} into equation (\ref{eq:Brillouin}) results in $M^{loose} = 0.03\,M^{loose}_S$. Thus, the induced magnetization of loose spins at room temperature amounts to only \SI{3}{\percent} of its saturation value. Therefore, at room temperature the contribution of loose spins is negligible. As a consequence, the induced perpendicular magnetization at RT is solely due to the antiferromagnetically coupled spins. This implies that the exchange constant of AF Mn$_2$Au thin films can be determined by measuring the induced magnetic moment at the highest available external field. However, at low temperatures the contribution of the loose spins is more significant and constitutes a substantial part of the measured magnetic signal.

\section{IV. EXPERIMENTAL RESULTS}

\begin{figure}[ht]
\includegraphics[width=0.45\textwidth]{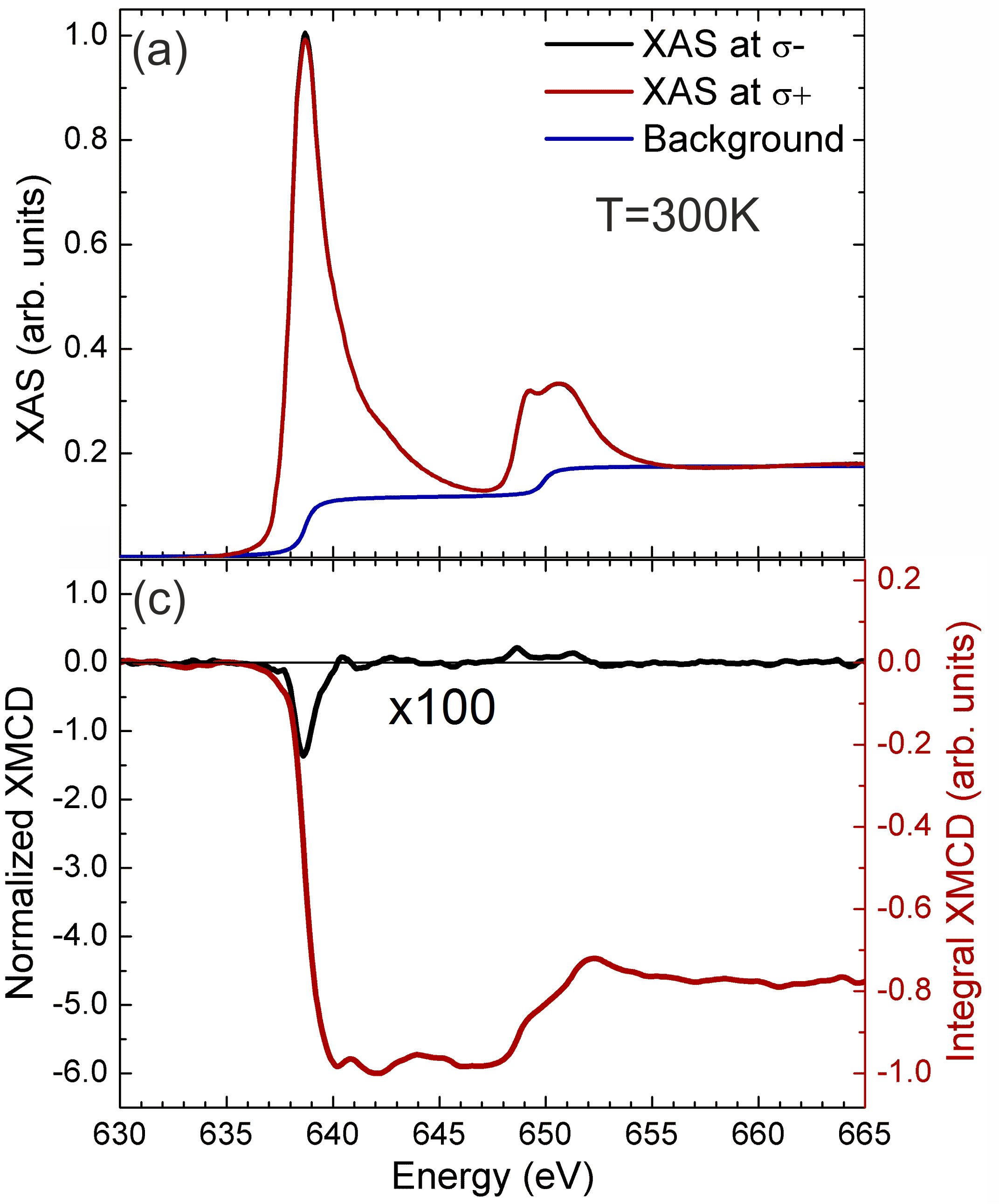}\hfill
\includegraphics[width=0.45\textwidth]{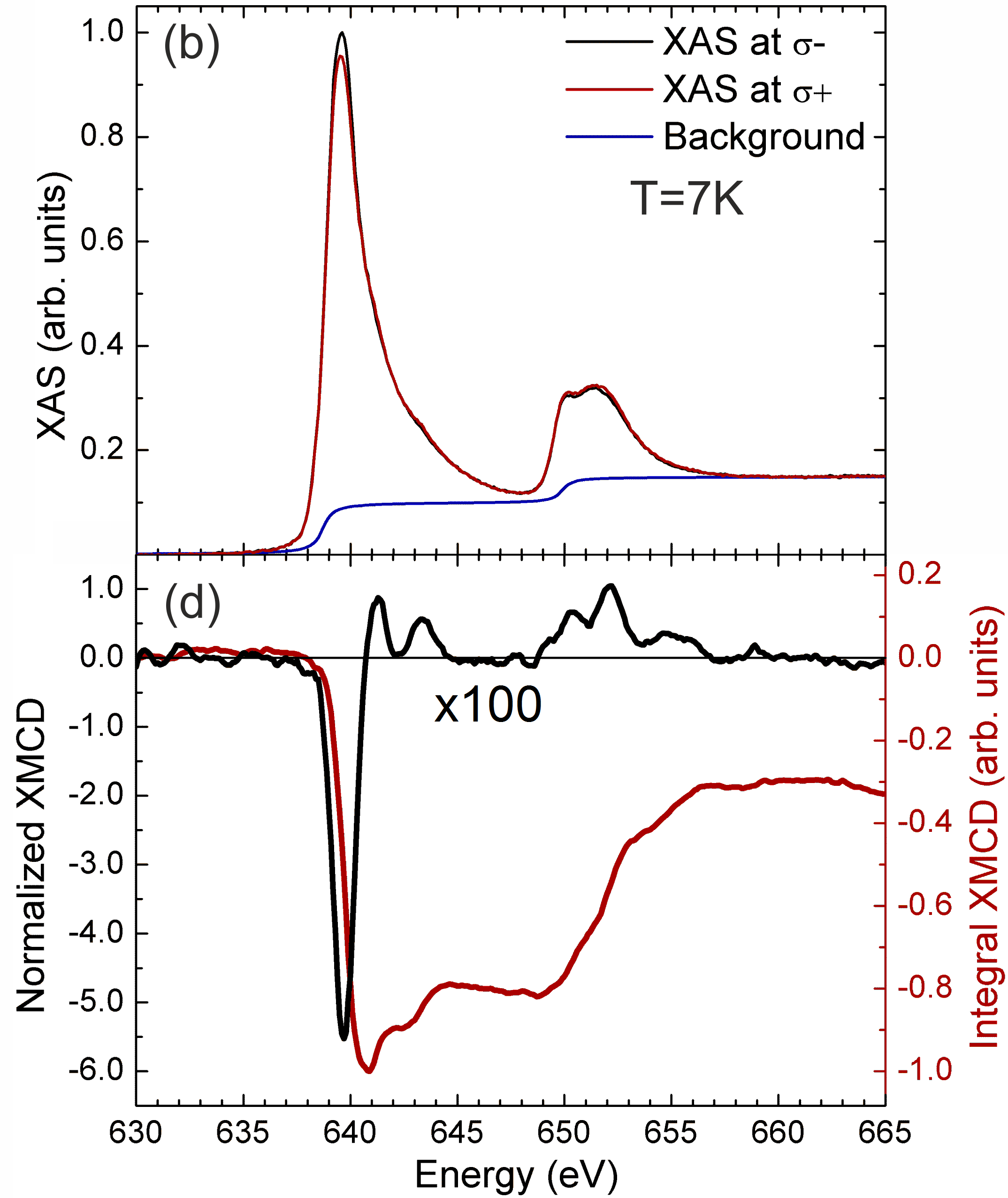}
\caption{XAS measured for the Mn$_2$Au film at \SI{300}{\kelvin} (a) and \SI{7}{\kelvin} (b), respectively, while applying \SI{8}{\tesla} out-of-plane magnetic field corresponding to circular positive (black) and negative (red) polarization. The blue line shows the contribution of the delocalized electrons to X-ray absorption. (c, d) XMCD calculated with the data presented in (a) and (b), respectively. The XMCD values are the difference between the corresponding XAS spectra multiplied by 100.}
\label{fig:XMCD_RT} 
\end{figure}

The X-ray absorption spectra (XAS) were measured for right and left circular X-ray polarization in a field of \SI{8}{\tesla} applied to the sample at \SI{300}{\kelvin} (Fig.\,\ref{fig:XMCD_RT}\,(a)) and at \SI{7}{\kelvin} (Fig.\,\ref{fig:XMCD_RT}\,(b)). The XMCD spectra resulting from the difference of two XAS at the respective temperatures and are shown in Fig.\,\ref{fig:XMCD_RT}\,(c, d).

The sum rules \cite{THO92, CAR93} were employed for determining the values of spin and orbital moment in the following form:
\begin{equation}
\begin{gathered}
r=2\int_{L_3+L_2} XAS_0(E)\;dE \\
p=\frac{C_{jj}}{P}\int_{L_3} XMCD(E)\;dE \\
q=\frac{C_{jj}}{P}\int_{L_3+L_2} XMCD(E)\;dE \\
\mu_S=N_d\,(6p-4q)/r \\
\mu_L=N_d\,4q/3r,
\label{eq:SumRules}
\end{gathered}
\end{equation}
where $N_d$ is the number of 3d holes in Mn, $C_{jj} = 1.4$ is the correction factor compensating for the $jj$-mixing \cite{GOE05, SCH05}, $P=0.77$ is the degree of circular polarization provided by the beamline, and $XAS_0$ is the average of $XAS$ measured with different circular polarizations.

\subsection{a. High temperature measurements}
At \SI{300}{\kelvin} and $\mu_0H$ = \SI{8}{\tesla} the sum rule analysis results in values for the spin and orbital moments of  $\mu_S$ = 0.06\,$\mu_B$ and $\mu_L$ = 0.02\,$\mu_B$. The $\mu_L$ value is about \SI{30}{\percent} of $\mu_S$. Moreover, the experimentally determined orbital moment is much larger than the theoretical value $\mu_L^{Mn}$ presented in the Section~III. Please note, that the high temperature results are mainly due to the antiferromagnetically coupled Mn moments in the AF films and not due to the loose spins.

\subsection{b. Low temperature measurements}
If the same experiment is performed at low temperatures, we observe a considerably larger value of the spin moment, which we attribute to the contribution of the loose Mn spins. For a magnetic field of \SI{8}{\tesla} and for a sample temperature of \SI{7}{\kelvin}, we obtain a value of 0.27~$\mu_B$ for the spin and 0.05~$\mu_B$ for the orbital moment.

\begin{figure*}[ht]
\includegraphics[height=5.25cm]{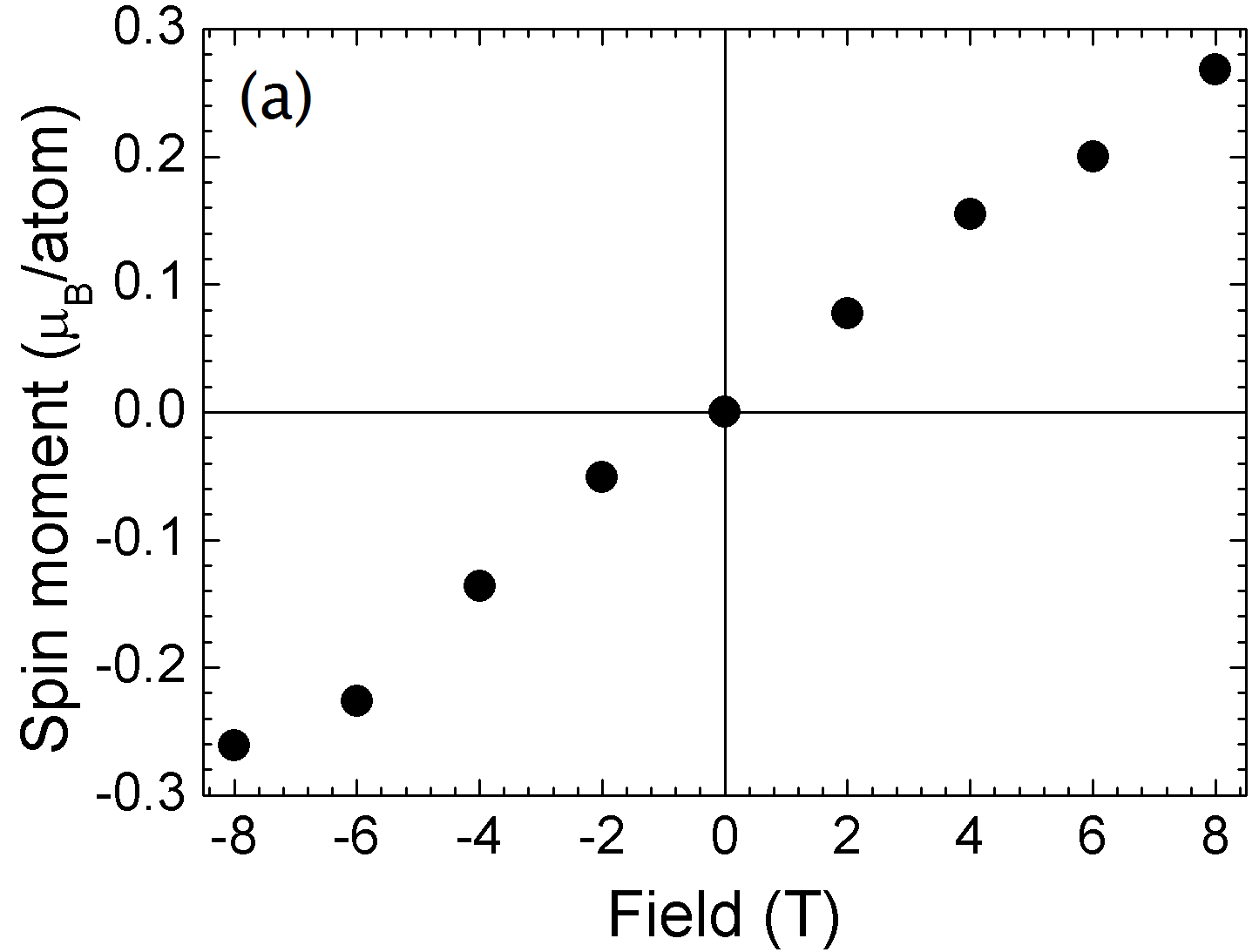} \hfill
\includegraphics[height=5.25cm]{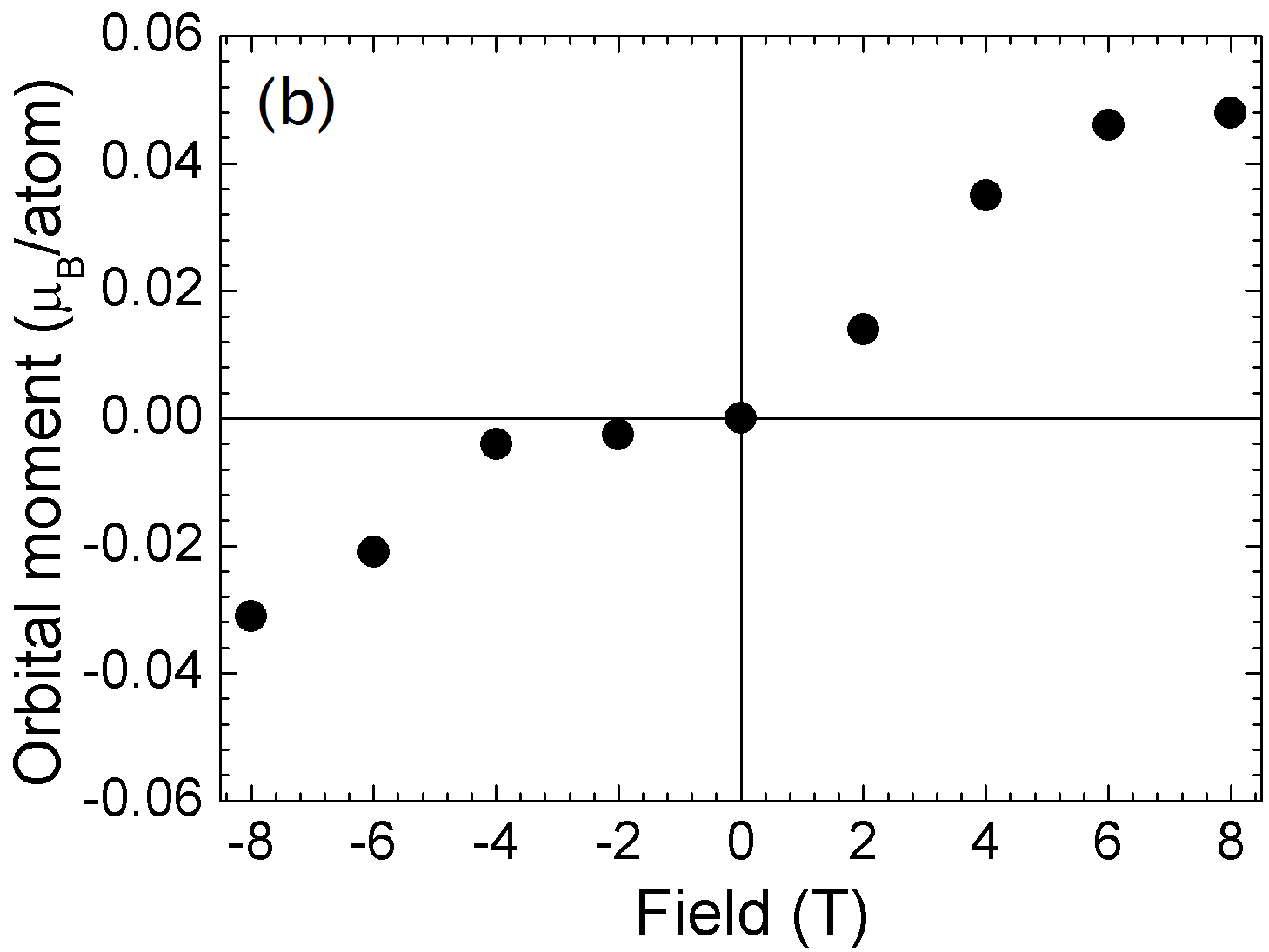}
\caption{(a) Spin magnetic moment and (b), orbital magnetic moment per Mn atom calculated from the corresponding XMCD spectra with sum-rules.}
\label{fig:XMCD_LT} 
\end{figure*}

The dependency of the moment values on the external field are shown in Fig.\,\ref{fig:XMCD_LT}. The ratio of orbital and spin moments in this case is about \SI{15}{\percent} independent on the magnetic field. The orbital to spin moment ratio is smaller at low temperatures, when the loose spin component dominates, than the value obtained at \SI{300}{\kelvin}. Thus, the larger $\mu_L$ in the high temperature case stems from the antiferromagnetically coupled moments.

\section{V. DISCUSSION}
\subsection {a. Perpendicular susceptibility of AF coupled spins}
\label{Discussion}
\begin{figure}[ht]
\includegraphics[scale=0.25]{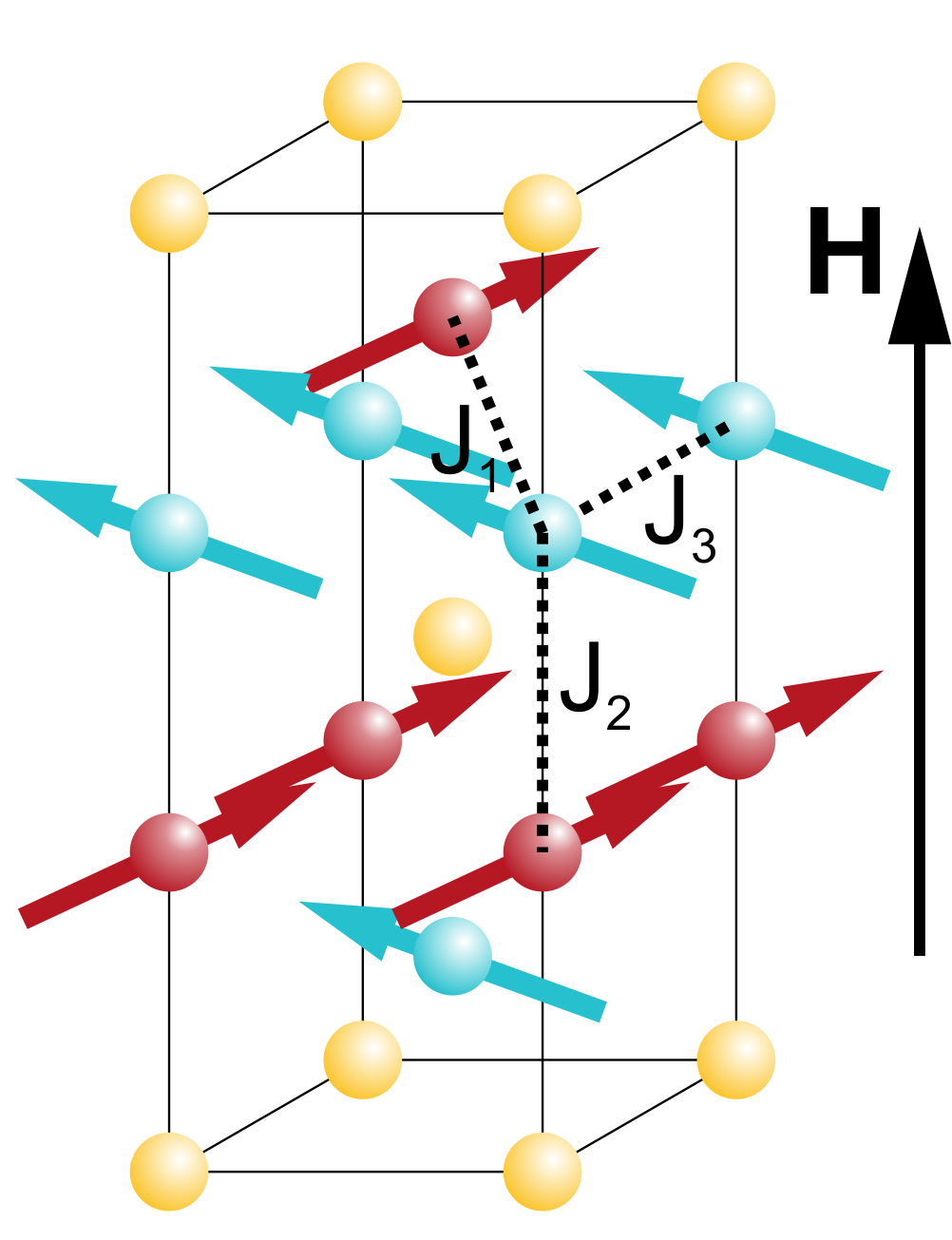}
\caption{The elementary cell of Mn$_2$Au, with Mn atoms drawn in blue and Au atoms in red, featuring the Mn spins reoriented in magnetic field applied along [001] direction. Three different exchange constants are specified following the notations from \cite{KHM08}.}
\label{fig:Mn2AuEC} 
\end{figure}

The model used for describing the perpendicular susceptibility of Mn$_2$Au employs three exchange constants, which describe the antiferromagnetic interaction between the neighboring Mn planes in Mn$_2$Au ($J_1$ and $J_2$) and the ferromagnetic interaction within one layer of Mn atoms ($J_3$) (Fig.\,\ref{fig:Mn2AuEC}). An external field applied along the [001] axis tilts the Mn moments such that the parallel alignment within the basal ab-planes is preserved. Thus, the exchange energy gain per Mn atom depends on the effective constant $J_{eff} = (4\, J_1 + J_2)/2$, which does not contain $J_3$.

The field induced changes in the Mn$_2$Au magnetic moment arrangement were analyzed with the help of the energy density functional, where $\epsilon$ is the energy per unit cell volume ($V_{u.c.}$) (Fig.\,\ref{fig:Mn2AuEC}). It includes exchange and Zeeman energy, magneto-crystalline anisotropy energy (MAE) and shape anisotropy:
\begin{equation}
\epsilon = \epsilon_{exchange} + \epsilon_{MAE} + \epsilon_{shape} + \epsilon_{Zeeman},
\label{eq:Energy}
\end{equation}
The exchange energy is written as:
\begin{equation}
\epsilon_{exchange} =-4J_{eff}\,\hat{\vec{e_1}}\cdot\hat{\vec{e_2}},
\label{eq:ExchangeApproximated}
\end{equation}
where $\hat{\vec{e_1}}$ and $\hat{\vec{e_2}}$ are the unit vectors oriented along the corresponding sublattice magnetizations. Within the approximation implying that the tilting angle ($\alpha$) of the Mn spins towards the film normal is small, Eq. (\ref{eq:ExchangeApproximated}) becomes:
\begin{equation}
\epsilon_{exchange} = -4J_{eff}\,\cos(\pi-2\alpha) \sim 8J_{eff}\,\alpha^2,
\label{eq:ExchangeApproximated2}
\end{equation}
where the constant term defining the zero-energy value is omitted. The magnetocrystalline anisotropy is given by:
\begin{equation}
\epsilon_{MAE} = 4K_{2\perp}\,\cos^2\alpha \sim 2K_{2\perp}\,\alpha^2,
\label{eq:MAEApproximated}
\end{equation}
where $K_{2\perp}$ is the Mn$_2$Au anisotropy coefficient equal to \SI{2.44}{\milli\electronvolt}/f.u. \cite{SHI10}. The shape anisotropy results from:

\begin{equation}
\epsilon_{shape} = \frac{\mu_0M_S^2}{2} \, V_{u.c.}\,\cos^2\alpha \sim \frac{\mu_0M_S^2}{2}V_{u.c.}\,\alpha^2,
\label{eq:ShapeApproximated}
\end{equation}
where $M_S$ is the Mn$_2$Au saturation magnetization.  

The total energy can be written as:
\begin{equation}
\epsilon = (8J_{eff}+2K_{2\perp}+\frac{\mu_0M_S^2}{2} \, V_{u.c.})\left(\frac{\mu_S}{\mu_S^{Mn}}\right)^2 - 4 (\mu_S + \mu_L) \mu_0 H,
\label{eq:EnergyApproximated}
\end{equation}
where $\alpha$ is replaced with $\mu_S/\mu_S^{Mn}$.

Note that both types of AF domains in a Mn$_2$Au(001) film, namely $\left[110\right]$ and $\left[1\bar{1}0\right]$ \cite{BAR15, SAP17}, produce the same contributions to the induced magnetic moment in an out-of-plane field. Minimizing (\ref{eq:EnergyApproximated}) with respect to the spin magnetic moment value, the following expression for $J_{eff}$ is obtained:
\begin{equation}
J_{eff} = \frac{\mu_S^{Mn}}{4}\mu_0H \frac{\mu_S^{Mn}}{\mu_S}-\left(\frac{K_{2\perp}}{2}+\frac{\mu_0M_S^2}{8} \, V_{u.c.}\right).
\label{eq:Jeff}
\end{equation}
Inserting values in Eq. (\ref{eq:Jeff}) from our experiments at \SI{300}{\kelvin} we find for J$_{eff}$ = \SI[separate-uncertainty=true]{24(5)}{\milli\electronvolt}. This is considerably less than the theoretically predicted value of \SI{90}{\milli\electronvolt} from Ref. \cite{KHM08} and the experimentally determined value of \SI{75}{\milli\electronvolt} reported earlier \cite{BAR13}. The reason for this discrepancy can be the reduction of the exchange constant value in the near-interface region of a thin film or the presence of loose spin clusters contributing to the susceptibility even at room temperature.

\begin{figure*}
\includegraphics[width=0.75\textwidth]{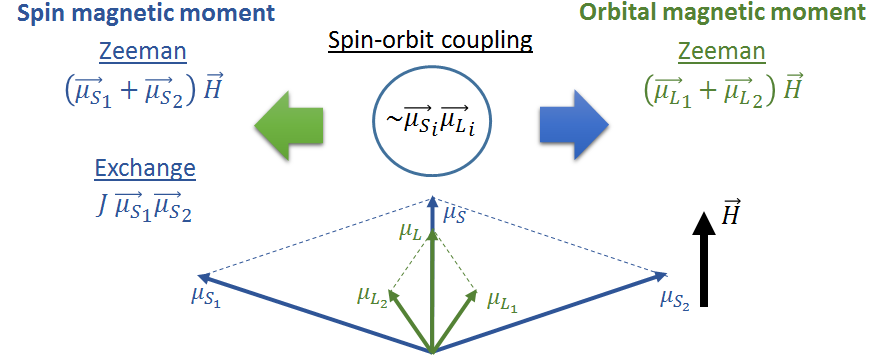}
\caption{The spin ($\mu_{S_1}$ and $\mu_{S_2}$) and the orbital ($\mu_{L_1}$ and $\mu_{L_2}$) magnetic moments of two Mn$_2$Au sublattices in external magnetic field. Spin-orbit coupling couples the spin and the orbital moment of the same sublattice.}
\label{fig:Model} 
\end{figure*}

The large measured values of $\mu_L$ can be interpreted with the following simple model (Fig.\,\ref{fig:Model}).
The spin magnetic moments of both sublattices are strongly coupled antiparallel to each other via exchange interaction, which the magnetic field counteracts on (left side of Fig.\,\ref{fig:Model}). The orbital magnetic moments are coupled to the spin moments via spin-orbit coupling (central part of the Fig.\,\ref{fig:Model}). It is worth mentioning that the model requires a sizable (not fully quenched) orbital moment of the magnetic atoms. The spin-orbit interaction constant is one order of magnitude smaller than the exchange energy \cite{DUR97}. Therefore, the external field rotates the orbital magnetic moments by a larger angle out of their equilibrium position than the spin moments. A similar non-collinear arrangement of spin and orbital magnetic moment has also been found in ferromagnets \cite{DUR97}. As a result, the projected perpendicular orbital moment is of the same magnitude as the projected spin moment. Hence, the non-collinear arrangement of spin and orbital moment explains the observed large orbital to spin moment ratio. Our experimental result strongly suggest that the exchange interaction exclusively acts on the spin moments and not on the orbital moments. 

\subsection {b. Perpendicular susceptibility of loose spins}
\begin{figure}[ht]
\includegraphics[width=7cm]{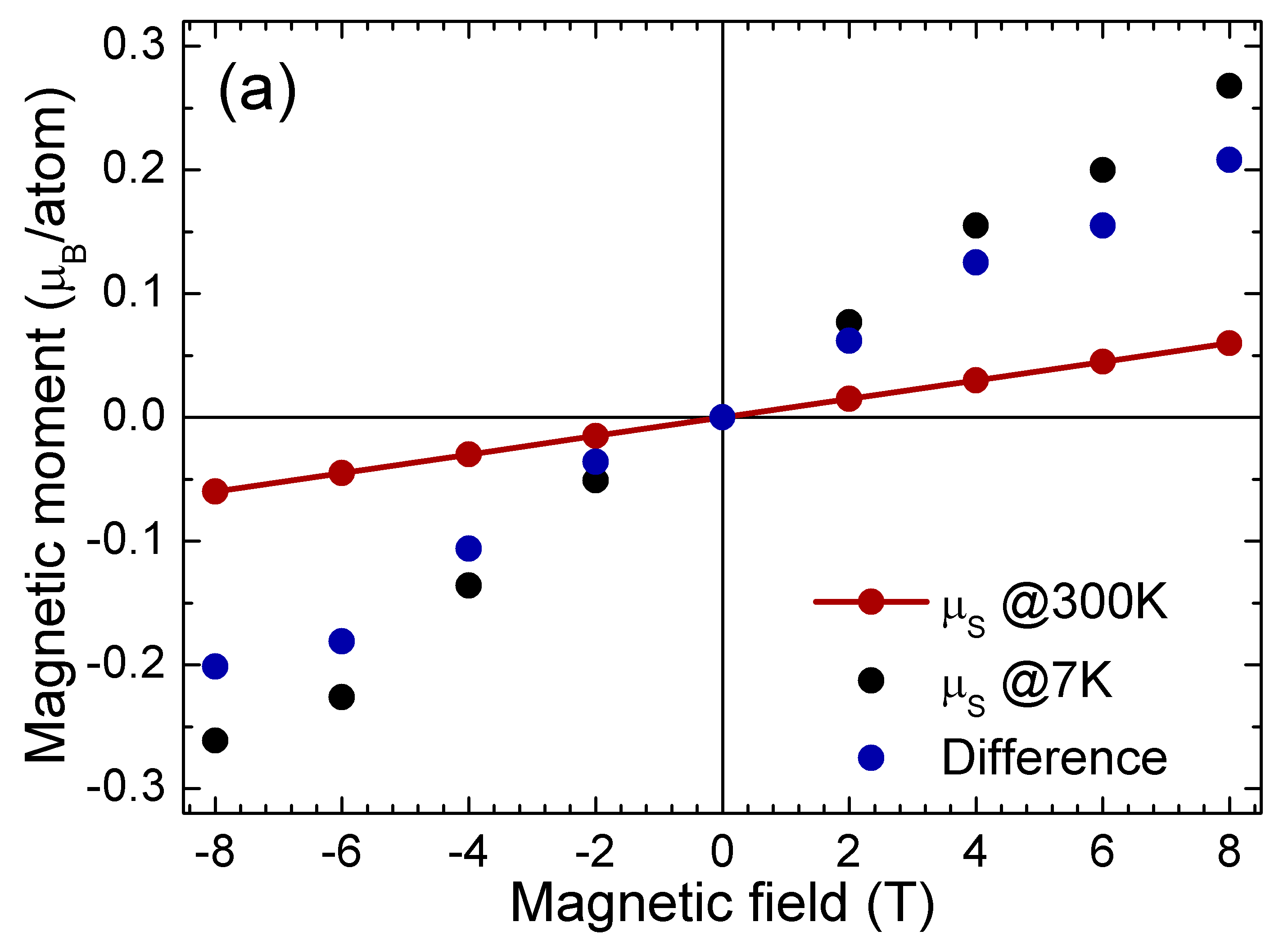}
\includegraphics[width=7cm]{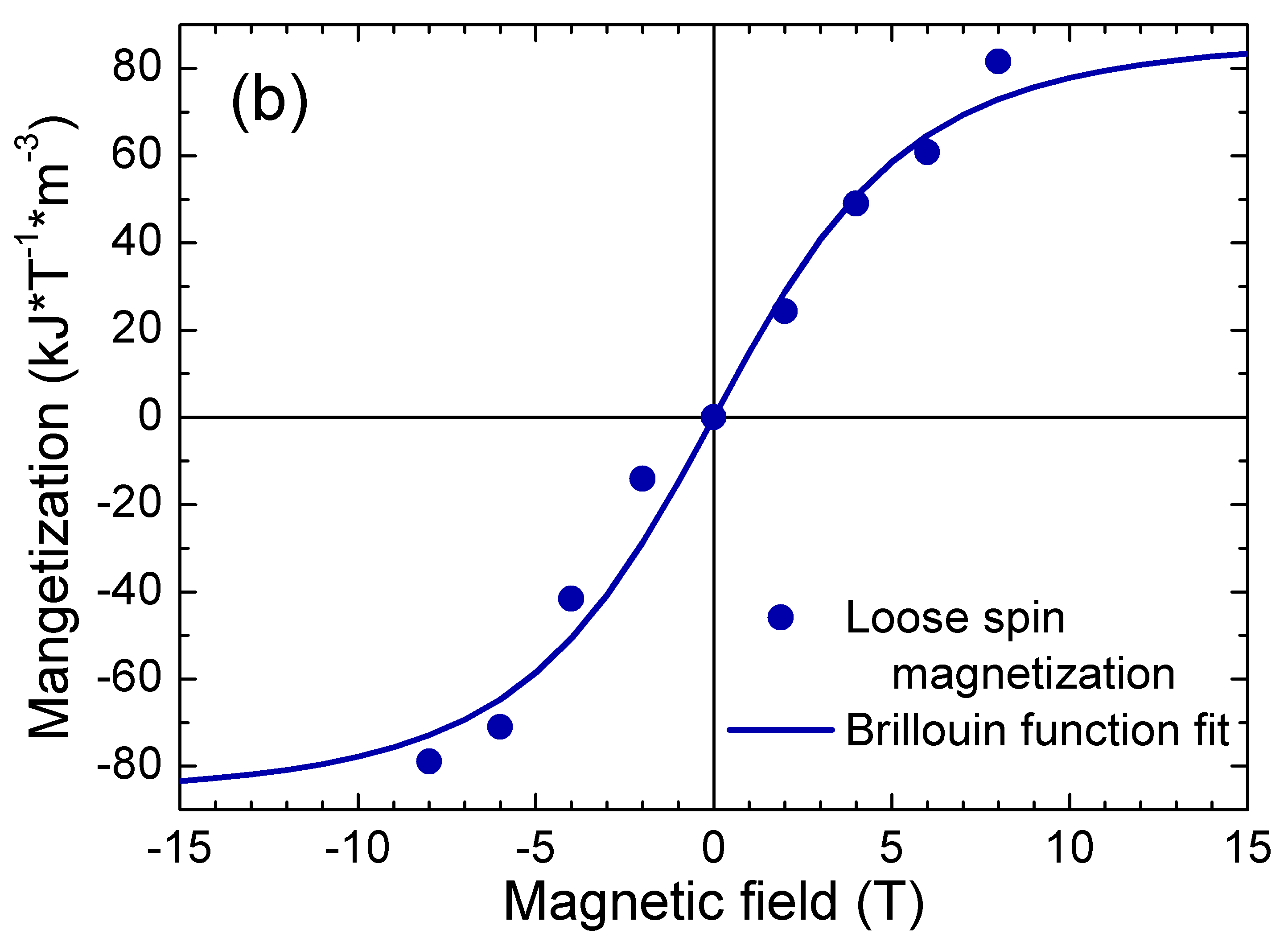}
\caption{(a) Field dependence of the spin magnetic moment at \SI{7}{\kelvin} and extrapolated values at \SI{300}{\kelvin}. The difference yields the contribution of loose spins at a temperature of \SI{7}{\kelvin} and assuming that their moments are equally distributed over all Mn atoms. (b) The magnetization of loose spins follows the Brillouin function providing the best fit to the experimental data with a saturation magnetization of \SI{86}{\kilo\joule\per\tesla \per\metre\tothe{3}}.}
\label{fig:LooseSpins} 
\end{figure}

The loose spins lead to an increase of the magnetic moments measured at low temperatures. The induced perpendicular moment of antiferromagnetically coupled atoms is independent on temperature. Thus, the room temperature value can be extrapolated to the lower magnetic fields and subtracted from the measured total moment at low temperature to find the loose spin contribution (Fig.\,\ref{fig:LooseSpins} (a)).

Since the sum rules provide the magnetic moment per atom, assuming a homogeneous spin distribution, for further analysis it is more convenient to recalculate this value in terms of magnetization using the Mn atom density of 4.2$\times$10$^{28}$ m$^{-3}$. The Brillouin function fit (Eq. (\ref{eq:Brillouin})) to the field dependent induced magnetization of the loose spins is presented in Fig.\,\ref{fig:LooseSpins} (b). The saturation magnetization $M^{loose}_{S}$ of the loose spins was the fitting parameter with the best fit value of $M^{loose}_{S}$ = \SI{86}{\kilo\joule\per\tesla \per\metre\tothe{3}}. This value corresponds to a number of loose spins per unit volume of \SI{2.6e26}{\metre\tothe{-3}} or \SI{6}{\percent} of all Mn spins in the $\sim$\SI{3}{\nano\metre} near-surface layer probed by X-ray absorption. This seems to be a reasonable value considering the determined grain size and the material stoichiometry (see Section II), with additional inter-site disorder.

\section{VI. CONCLUSIONS}
Using XMCD, the perpendicular susceptibility of a \SI{10}{\nano\metre} thick Mn$_2$Au layer in a high magnetic field has been determined. From the perpendicular susceptibility, we find an effective exchange constant of \SI[separate-uncertainty=true]{24(5)}{\milli\electronvolt}. The measured orbital magnetic moment is surprisingly large and amounts to about \SI{30}{\percent} of the spin moment. A model considering exclusively spin-spin exchange coupling in combination with a weaker spin-orbit coupling is proposed for explaining this phenomenon. Complementary measurements at low temperatures provided the concentration of loose spins in the near-surface region, which for our sample turns out to be \SI{6}{\percent} of the total amount of Mn spins in Mn$_2$Au. The presented procedures for determining effective exchange constants, magnetic moments, and loose spin concentrations via XMCD measurements can be used for characterizing a wide range of AF materials, which is important for identifying novel materials suitable for AF spintronics applications.

\begin{acknowledgments}
The research was financially supported by the German Research Foundation (Deutsche Forschungsgemeinschaft) through the Transregional Collaborative Research Center 173 ”Spin+X”, Project A05. A.\,A.\,S. also wishes to acknowledge the fellowship of the MAINZ Graduate School of Excellence. We thankfully acknowledge the financial support from HZB. The authors also gratefully acknowledge the computing time granted on the supercomputer Mogon at Johannes Gutenberg University Mainz (hpc.uni-mainz.de). Financial support for developing and building the PM2-VEKMAG beamline and VEKMAG end station was provided by HZB and BMBF (05K10PC2, 05K10WR1, 05K10KE1), respectively. Steffen Rudorff is acknowledged for technical support.
\end{acknowledgments}


\bibliography{apssamp}
\end{document}